\documentclass[a4paper]{spie}  

\usepackage{amsmath,amsfonts,amssymb}
\usepackage{graphicx}
\graphicspath{{./}{Figures/}}
\usepackage[colorlinks=true, allcolors=blue]{hyperref}
\usepackage{comment}
\usepackage{soul}
\usepackage{lineno}

\def\LB{\textit{LiteBIRD}}

\title{Testbed preparation of a small prototype polarization modulator for LiteBIRD low-frequency telescope}

\author[a]{Thuong D. Hoang}
\author[a]{Tomotake Matsumura}
\author[b]{Ryota Takaku}
\author[a]{Takashi Hasebe}
\author[a]{Tommaso Ghigna}
\author[a]{Nobuhiko Katayama}
\author[c]{Yuki Sakurai}
\author[c]{Kunimoto Komatsu}
\author[d]{Teruhito Iida}
\author[e]{Yurika Hoshino}
\author[e]{Shinya Sugiyama}
\author[c]{Hirokazu Ishino}
\author[ ]{for the \LB\ collaboration.}

\affil[a]{Kavli IPMU (WPI), UTIAS, The University of Tokyo, Kashiwa, Chiba 277-8583, Japan}
\affil[b]{Department of Physics, The University of Tokyo, Tokyo, Japan}
\affil[c]{Okayama University, 3-1-1, Tsushimanaka, Kita-ku, Okayama, Okayama, 700-8530, Japan}
\affil[d]{ispace inc., Tokyo Japan}
\affil[e]{Saitama University, 255 Shimookubo, Sakura-ku, Saitama, 338-8570, Japan}

\authorinfo{Further author information: (Send correspondence to Thuong D. Hoang)\\Thuong D. Hoang: E-mail: thuong.hoang@ipmu.jp}

\begin{document}
\maketitle

\begin{abstract}
\LB\ is the Cosmic Microwave Background (CMB) radiation polarization satellite mission led by ISAS/JAXA. The main scientific goal is to search for primordial gravitational wave signals generated from the inflation epoch of the Universe. \LB\ telescopes employ polarization modulation units (PMU) using continuously rotating half-wave plates (HWP). The PMU is a crucial component to reach unprecedented sensitivity by mitigating systematic effects, including 1/f noise. We have developed a 1/10 scale prototype PMU of the \LB\ LFT, which has a 5-layer achromatic HWP and a diameter of 50 mm, spanning the observational frequency range of 34-161~GHz. The HWP is mounted on a superconducting magnetic bearing (SMB) as a rotor and levitated by a high-temperature superconductor as a stator. In this study, the entire PMU system is cooled down to 10~K in the cryostat chamber by a 4-K Gifford-McMahon (GM) cooler. We propagate an incident coherent millimeter-wave polarized signal throughout the rotating HWP and detect the modulated signal. We study the modulated optical signal and any rotational synchronous signals from the rotation mechanism. We describe the testbed system and the preliminary data acquired from this setup. This testbed is built to integrate the broadband HWP PMU and evaluate the potential systematic effects in the optical data. This way, we can plan with a full-scale model, which takes a long time for preparation and testing.
\end{abstract}


\section{INTRODUCTION}
\label{sec:intro} 
\LB\ is a satellite mission to measure the cosmic microwave background (CMB) over the full sky at a large angular scale, searching for the inflationary B-mode signal. The \LB's\ main scientific goal is to achieve the sensitivity in the tensor-to-scalar ratio, $r<0.001$ ~\cite{LB_PTEP_2022}. The \LB's\ focal plane accommodates $\sim5000$ multi-chroic polarized Transition Edge-sensor (TES) bolometers covering a wide range of frequencies from 34 to 448~GHz. \LB\ payload module (PLM) consists of Low-Frequency Telescope (LFT), Mid-Frequency Telescope (MHT), and High-Frequency Telescope (HFT). These telescopes operate at a cryogenic temperature of 5~K. Each telescope deploys a polarization modulation unit (PMU) which is the first optical element of the satellite. The PMU contains a continuously rotating broadband half-wave plate (HWP) to modulate the incident CMB polarization signal.

Measurement of the B-mode polarization signal at large angular scales can be contaminated by $\rm 1/f$ noise\cite{Kusaka_2014, Hill2020}. The temperature-to-polarization leakage also contaminates it by the instrumentally induced systematic effects, e.g. beam shapes, band-passes filter mismatch \cite{Hoang_2017}, and different gains among detectors. A HWP is the key instrumental element in mitigating systematic effects. The precise measurement of the CMB polarization signal requires accurate characterizations and models of all satellite subsystems, including the first optical element, PMU.

This study is a part of the development program of the \LB\ LFT PMU~\cite{Tomo_2016, Yuki_2020, Komatsu_2020, Takaku_2020, Komatsu_2021}. We made a $\rm 1/10$ scale prototype PMU, which contains a five-layer achromatic HWP (AHWP). A single wave plate is a sapphire disc with a diameter of 50~mm. The first and fifth layers include anti-reflective sub-wavelength structures. This small prototype PMU contains all the sub-components, which are to be scaled to a full-scale LFT PMU model, e.g. a cryogenic holder mechanism, rotation mechanism, superconducting magnetic bearing (SMB), encoder, readout monitoring, and drive electronics. Therefore, this prototype PMU system is a valuable development model for understanding the potential malfunctions, systematic effects, and unexpected features we anticipate in an upcoming full-scale model, reducing the risk involved in the full-scale development program. Examples are the HWP rotor vibrations, the HWP position angle reconstruction, the heat dissipation, and the realistic modulated signal from the HWP rotation.

In this paper, we describe a small prototype PMU, which is placed inside a 4-K Gifford-McMahon (GM) cryostat operating at 10~K. A part of the development status and the components of the small prototype PMU have been previously reported in Komatsu et al.\cite{Komatsu_2020}. We describe the progress and some of the sub-components, which are not detailed in the previous report.
 
The paper is structured as follows: Section \ref{sec:exp} presents the experimental setup of the small prototype PMU inside a cryostat. Section \ref{sec:result} shows the preliminary results of the angle reconstruction, the heat dissipation estimation, and the modulated signal of the continuously rotating AHWP. Section \ref{sec:discus} discusses future improvements, the misalignment of the optical setup, and the impact of the inhomogeneous magnetic fields on the focal plane. Finally, we summarize the proceeding of this paper in Section \ref{sec:conclu}.

\section{EXPERIMENT SETUP}
\label{sec:exp}
\begin{figure}[ht]
\begin{center}
    \begin{tabular}{c}
        \includegraphics[width=0.9\textwidth]{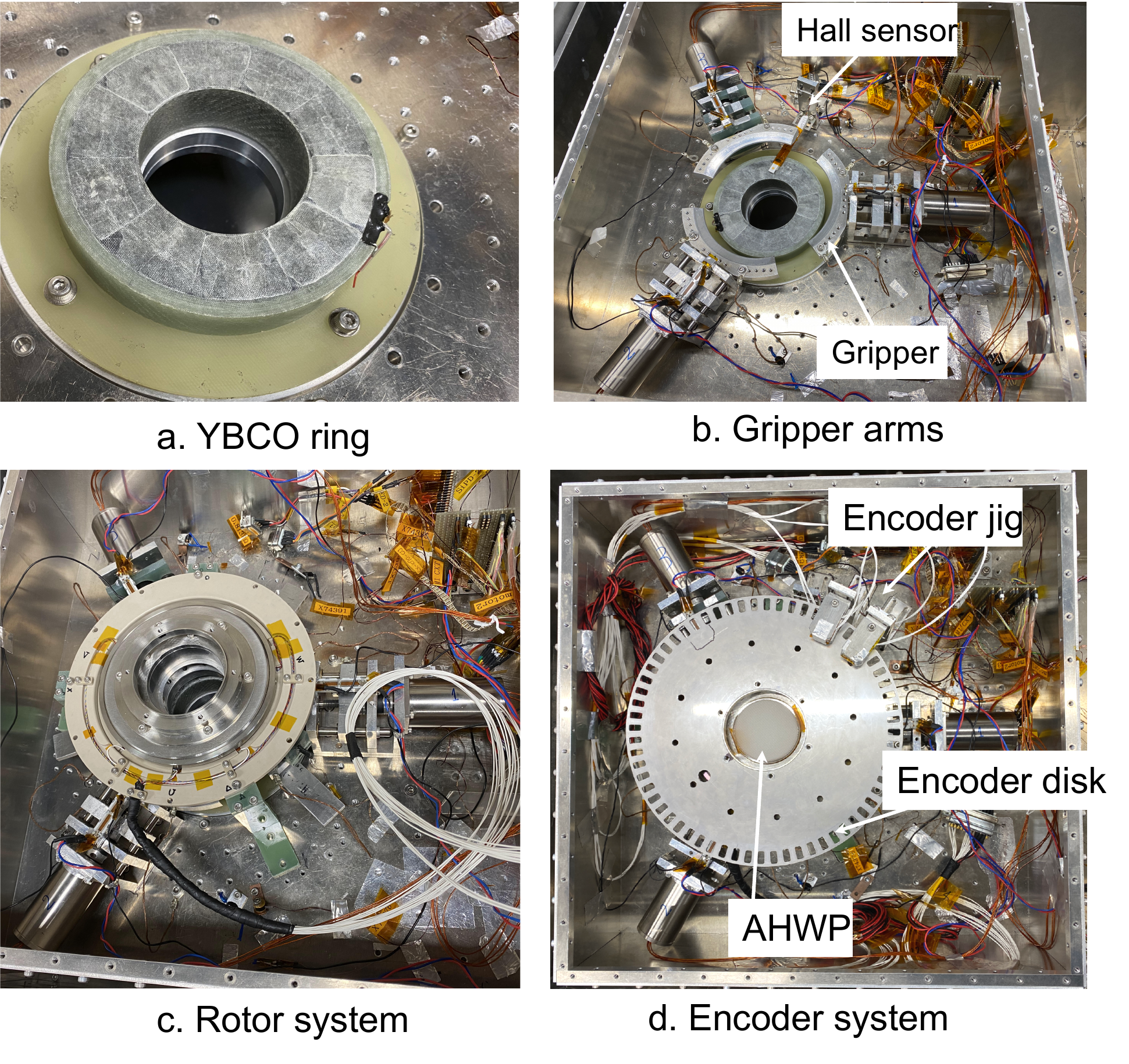}
    \end{tabular}
\end{center}
\caption[Instrument]{ \label{fig:instru} Photographs of the YBCO ring, gripper arms, the Hall sensor location, the installation of the rotor system, the encoder subsystem and the AHWP.}
\end{figure}

In this section, we describe the sub-components of the PMU in detail: the rotational mechanism, the gripper, the encoder, and the AHWP. Figure~\ref{fig:instru} shows the photograph of the sub-components. Then we describe an assembled setup.

\subsection{Cryogenic rotation mechanism}
We aim to maintain a stable rotation of the AHWP at 5~K. The cryogenic rotation mechanism employs a superconducting magnetic bearing (SMB) \cite{Tomo_2016}. The SMB consists of a high-temperature superconductor (HTS) YBCO and a NdFeB permanent magnet ring. The ten segments of YBCO bulk superconductors are arranged in a ring shape with an outer diameter, an inner diameter, and a height of 95~mm, 55~mm, and 20~mm, respectively. The NdFeB permanent magnet ring has an outer diameter of 85~mm, an inner diameter of 65~mm, and a height of 10~mm. At temperatures below 20~K, the frozen magnetic fields of the HTS will levitate the permanent magnet ring as a rotor. Therefore, the SMB with no mechanical contact can avoid the heat dissipation from the physical friction. However, we anticipate the energy loss due to the magnetic interaction, e.g. eddy current and hysteresis loss in the rotor and stator components~\cite{Yuki_2020}. One of the magnetic field sources is the rotor magnet. Any inhomogeneity of the ring magnet can produce the ac time-varying magnetic field for the stator components. We installed a Hall sensor (BHT921) and monitored its magnetic field inhomogeneity. The SMB can levitate the rotor magnet and thus the HWP. We must, however, drive the rotor and maintain its rotation at a constant rotational frequency. We employ a custom motor and drive electronics developed by Tamagawa Seiki. This AC driver motor outputs a three-phase signal with changing frequency depending on the feedback of the rotor speed.

\subsection{Cryogenic holder mechanism}
We use three cryogenic holder mechanisms separated by 120~degrees around the rotor. Each of them is controlled by a cryogenic stepping motor with a variable resistant slide volume, which implements the linear movement triggered by the stepping motor. This holder mechanism holds the rotor until the YBCO cools down below its critical temperature and functions as an SMB. Also, this holder serves as a conductive thermal path to cool the rotor because the thermal path is only through radiative heat exchange once the rotor levitates.

\subsection{Encoder system}
We measure the HWP position angle by an optical encoder. It consists of an encoder chopper disk on the rotor and a set of LED and silicon photodiode. We prepare three sets of LED (L9337-01) as emitters and SiPD (S2386-18L) as receivers. Each pair is placed face-to-face. The encoder disk has 64 slots for the relative angle and one for the absolute angle within one revolution. Two pairs of the LED-SiPD have represented phases A and B for the relative angle, while the other pair is used for the absolute angle named phase Z. Figure~\ref{fig:instru} shows the photograph of the encoder jigs and the encoder disk.

\subsection{Achromatic Half-Wave Plate}

The AHWP is a five-layer sapphire stack with a diameter of 50~mm in diameter\cite{}. The detailed description of this AHWP can be found in  \cite{Komatsu_2020, Komatsu_2021}. The thickness of the single a-cut HWP is chosen based on the center frequency of the LFT at 97.5 GHz. Sapphire HWP can reflect about half of the incident signal due to its high refractive index. Therefore, the first and fifth layers include a moth-eye anti-reflective sub-wavelength structure (SWS) by laser machining. This method performs $ >90 \%$ of transmittance over the frequency range from 43 to 161~GHz. The details can be found in \cite{Takaku_2020}.

\subsection{PMU System in the Cryostat}
\label{sub:exp}

\begin{figure} [ht]
\begin{center}
    \begin{tabular}{c}
        \includegraphics[width=1\textwidth]{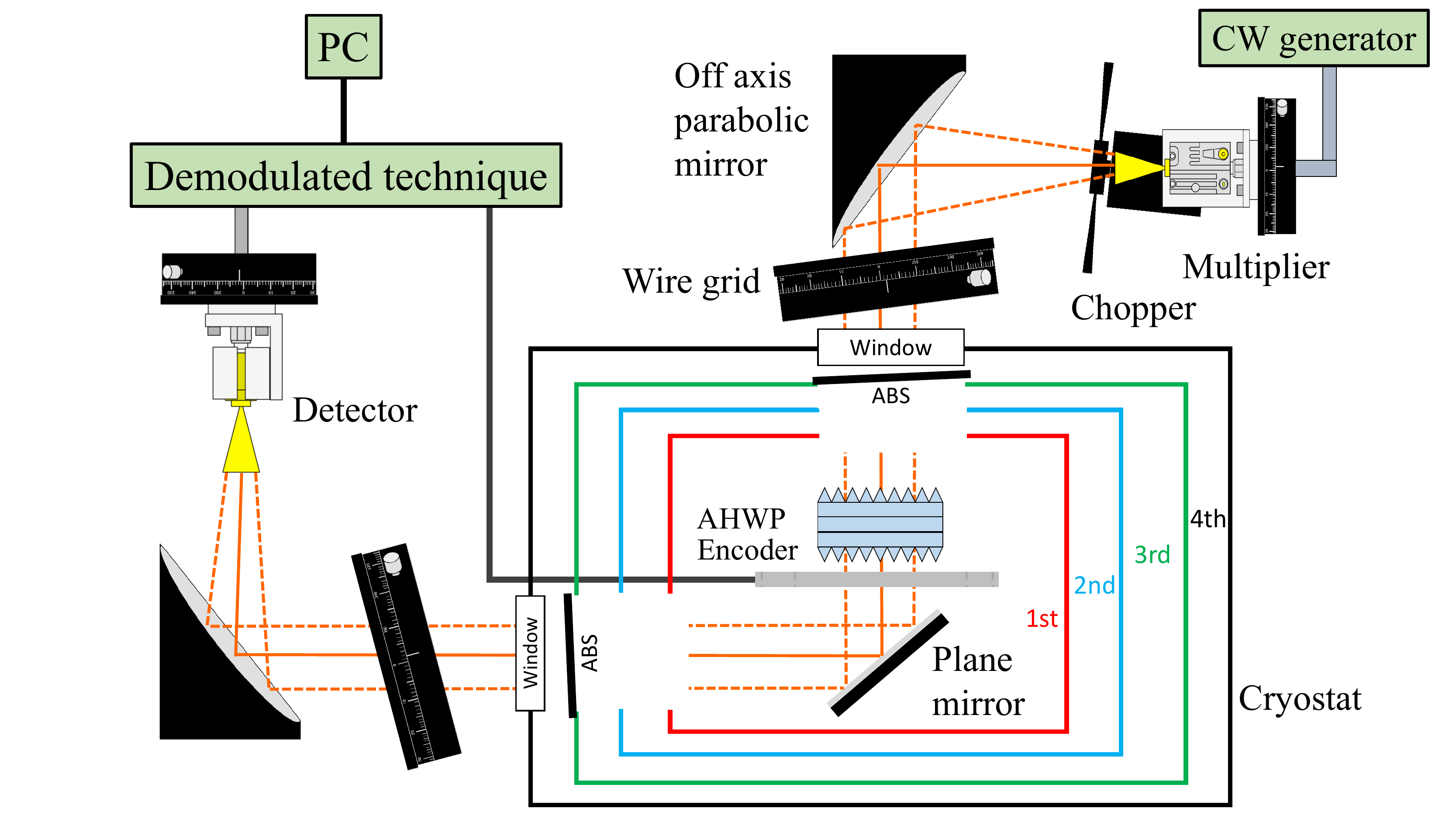}
    \end{tabular}
\end{center}
\caption[Optical measurement]{\label{fig:optic} 
The set up for the optical measurements. We use a GM cryocooler cryostat, off-axis parabolic mirrors, wire grids, and the chopper. The vacuum windows are UHMWPE plates. ABS stands for Acrylonitrile Butadiene Styrene.}
\end{figure}

Figure~\ref{fig:optic} shows the diagram of the experimental setup. A cryostat has two open windows with a diameter of 100~mm. A UHMWPE window is used to transmit the millimeter-wave source from the outside of the cryostat. We use a 2~mm thick Acrylonitrile butadiene styrene (ABS) plate as an infrared filter. The attenuated signal goes through AHWP mounted on the rotor of the rotational mechanism. Then, a 45~degree plane mirror reflects the signal inside the cryostat. Finally, a diode detector receives the output signal outside of the cryostat. 
At room temperature, a mechanical chopper chops the input millimeter-wave signal. In order to obtain the polarized signal, two-wire grids are placed to align the polarization angle.

The PMU system is cooled down to below 10~K using a 4-K GM cryocooler. After the cryogenic holder mechanisms are fully opened, the rotor levitates by the SMB mechanism. The motor applies the driver torque to spin the rotor at a constant speed. We then provide the incident signal through the rotating AHWP into the cryostat and detect the output modulated signal. Finally, we monitor the temperature of the PMU using thermometers.

As the description of the AHWP formalism \cite{Tomo_thesis_2006, Komatsu_2020,  Komatsu_2021}, the output modulated signal can be fitted using the equation
\begin{equation}
    I_{out}(\nu, t)=A_0(\nu) + \sum_{n=1}^{8} A_n (\nu) \cos \left( n \omega_{hwp} t + n \phi_n \right),
    \label{eq:modulate}
\end{equation}
where $ A_0$ is a constant amplitude, $ A_n$ are the amplitudes, $ \omega_{hwp}$ and $ \phi_n$ are the HWP modulated angle and the phase, respectively. We limit the harmonic frequency up to the $ 8^{th}$ in Equation \ref{eq:modulate} to study the synchronous signal.

\section{RESULTS}
\label{sec:result}

In this section, we present the preliminary results obtained from the experiment. Once the PMU system cooled down below 10~K, we spin the rotor using the electromagnetic drive mechanism and measure its rotation by the encoder signal. Then, we present the result of the angle reconstruction and the spin-down measurement. We also show the modulated signal, including the rotating AHWP synchronous signal. 

\subsection{Angle reconstruction}
\label{subsec:angle_res}
\begin{figure}[h]
\begin{center}
    \begin{tabular}{cc} 
        \includegraphics[width=0.48\textwidth]{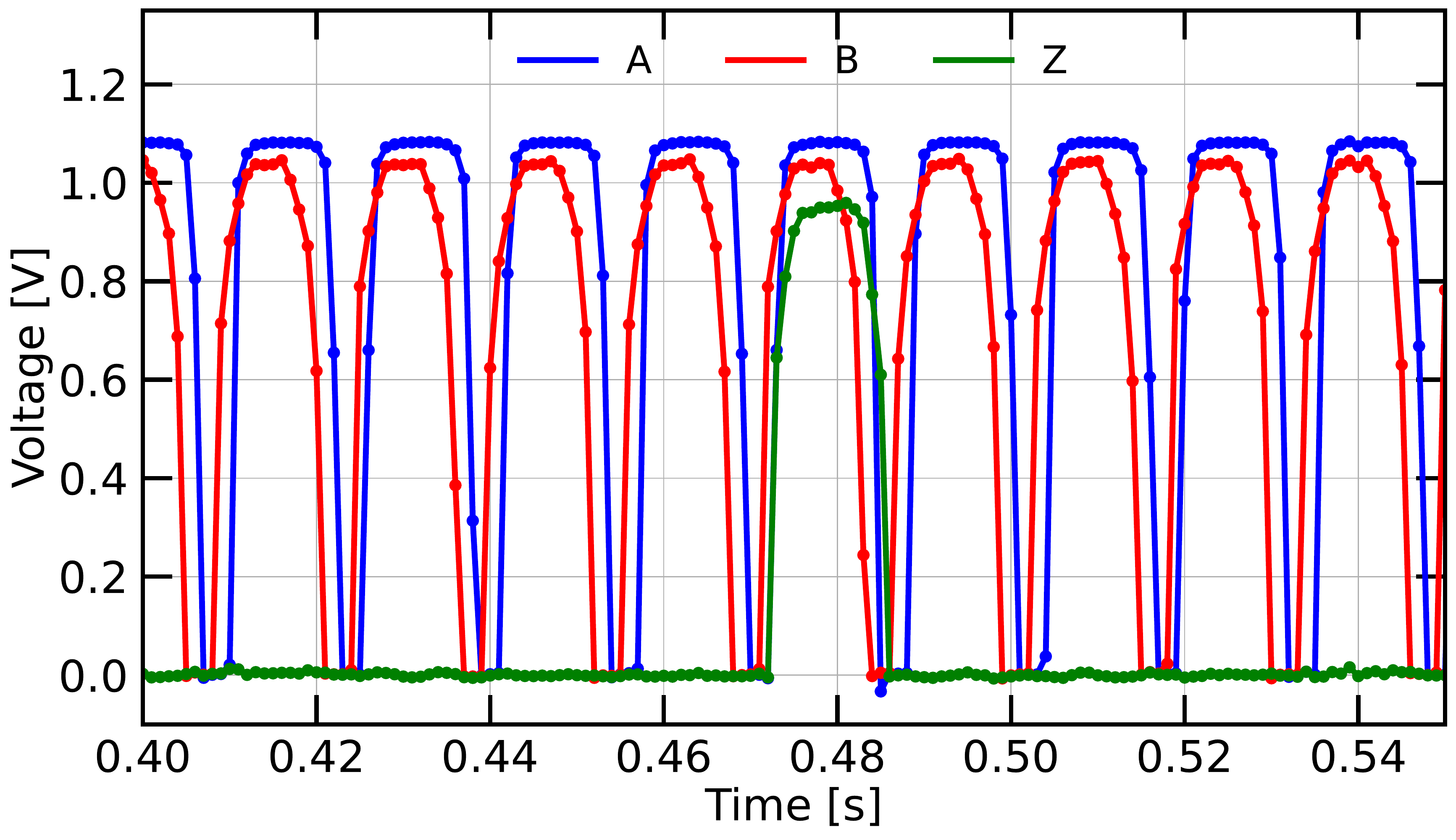}
        \includegraphics[width=0.5\textwidth]{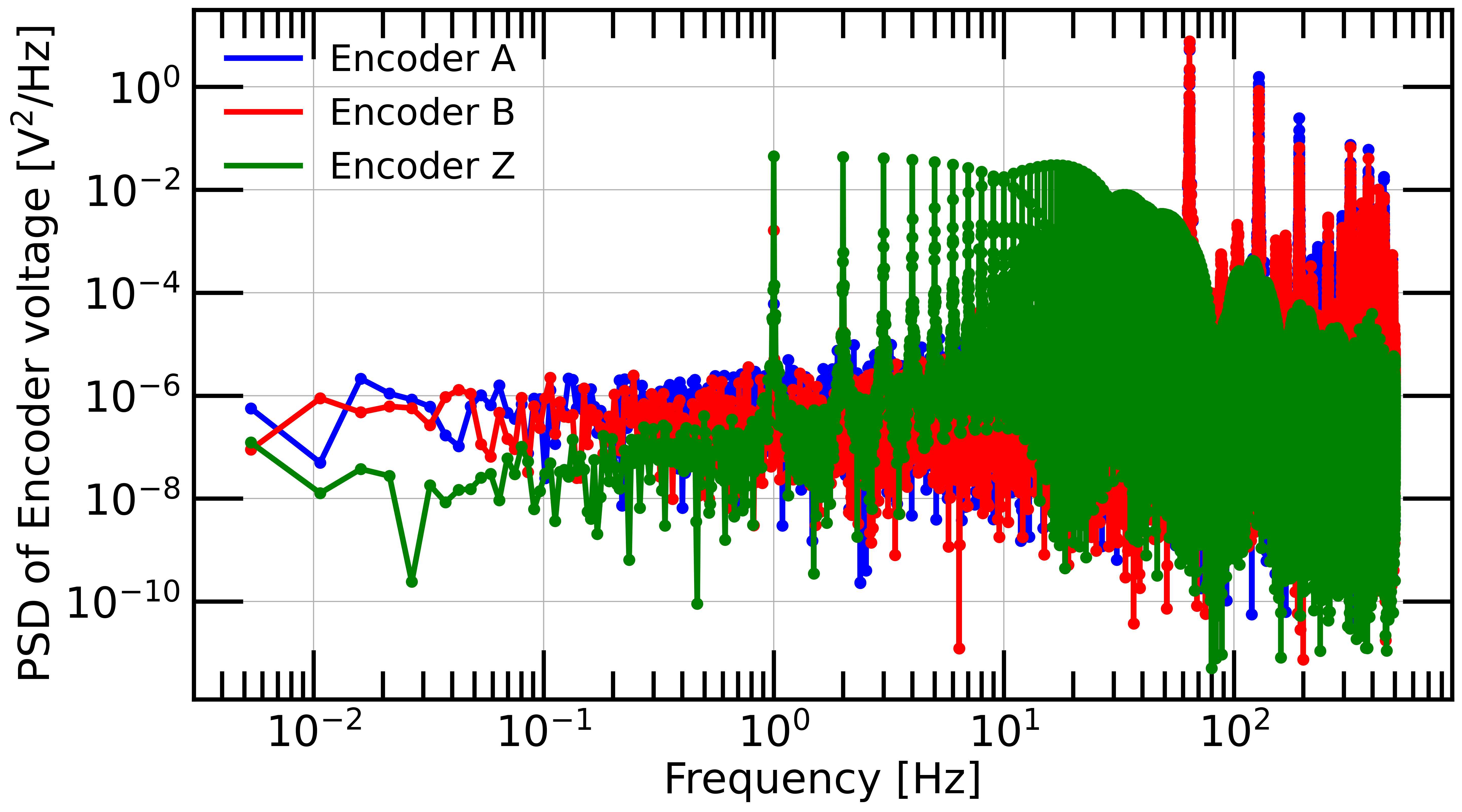}
    \end{tabular}
\end{center}
\caption[Encoder signal]{\label{fig:encoder} The encoder signal in time domain (left) and frequency domain (right) when spinning the rotor at 1Hz.}
\end{figure}

The left panel of Figure~\ref{fig:encoder} shows the raw encoder data when the PMU spins at 1~Hz. On the right panel, the power spectral density (PSD) of the encoder Z signal shows the dominant peak at 1~Hz and its harmonic oscillations of the PMU rotational frequency. The PSD of encoder A and B signals show the dominant peaks at the expected frequency of 64 Hz, corresponding to the 64 slots of the encoder disk.

\begin{figure}[h]
    \includegraphics[width=1\textwidth]{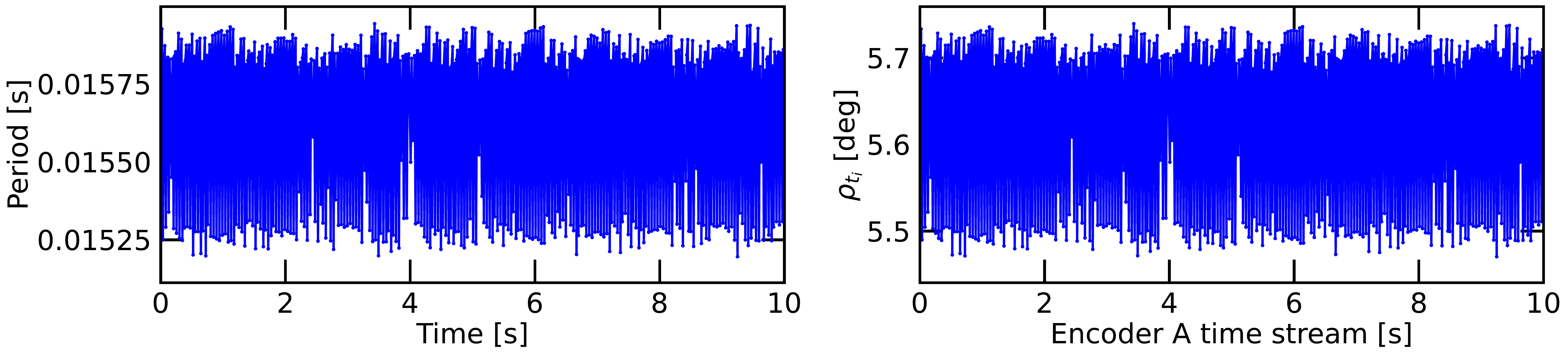}
\caption[Angle reconstruct]{\label{fig:ang_recontr} The period $\Delta t_i$ (left) and the HWP angle position reconstruction $\rho_{t_i}$ (right) for the encoder A signal in case of $f_{hwp} = 1 \ Hz$. The time in the horizontal axis is the $ t_i$ in the time stream.} 
\end{figure}

Figure~\ref{fig:ang_recontr} shows the period $\Delta t_i$ and the HWP angle position reconstruction $\rho_{t_i}$ from the encoder-A signal. We rotated the HWP at 1 Hz, the encoder disk has 64 slots as mentioned above, thus each slot is equivalent to an angle of $360/64\simeq5.6$~degrees. We estimated the uncertainty of the HWP angle position reconstruction $\sigma_\rho \sim 0.1$~degrees. 

\subsection{Spin-down measurement}
We conducted spin-down measurements to estimate the heat dissipation due to the rotational loss. Spin-down measurements help to estimate the energy loss from the contactless rotational mechanism. We set the speed of the rotor at a constant frequency and then let it freely decelerate. The angular deceleration can be expressed as a function of the HWP rotational frequency $f_{hwp}$. \cite{hanany_2003}
\begin{equation}
   \alpha =   2 \pi \dfrac{df_{hwp}}{dt} = a_0 + 2\pi a_1 f_{hwp}.
  \label{eq:spindown}
\end{equation}
The coefficient $a_0$ represents the contribution of the hysteresis loss due to the deceleration of the rotor. The coefficient $a_1$ determines the amount of the eddy current loss. The solution of the differential equation \ref{eq:spindown} has an exponential form
\begin{equation}
    f_{hwp} \sim \dfrac{-a_0}{2 \pi a_1} + \dfrac{1}{2 \pi a_1} e^{-a_1 (t + c)},
    \label{eq:sol_spindown}
\end{equation}
where $ c$ is a constant represents the starting point of the fitting time.
\begin{figure} [ht]
\begin{center}
    \begin{tabular}{c c}
        \includegraphics[width=0.49\textwidth]{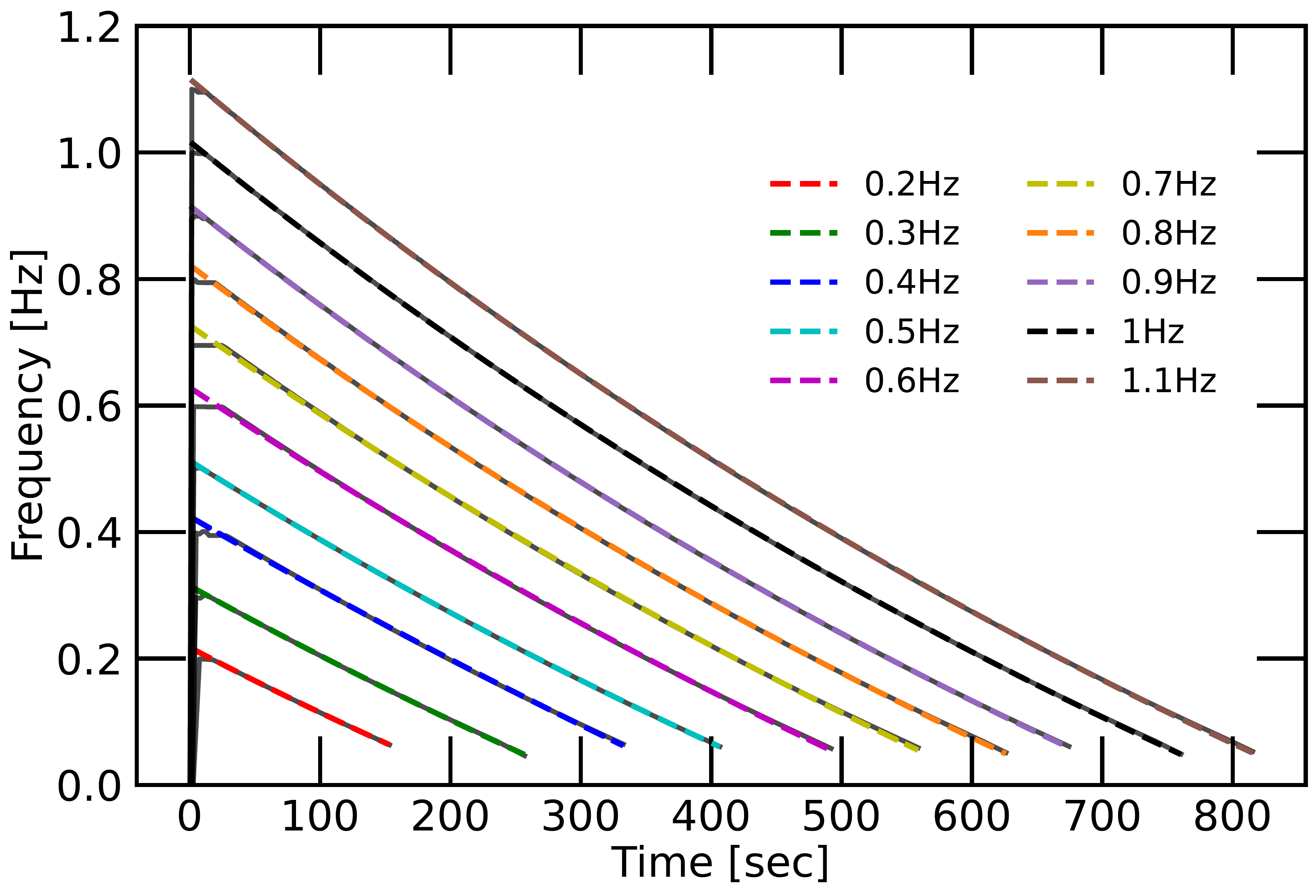}
        \includegraphics[width=0.5\textwidth]{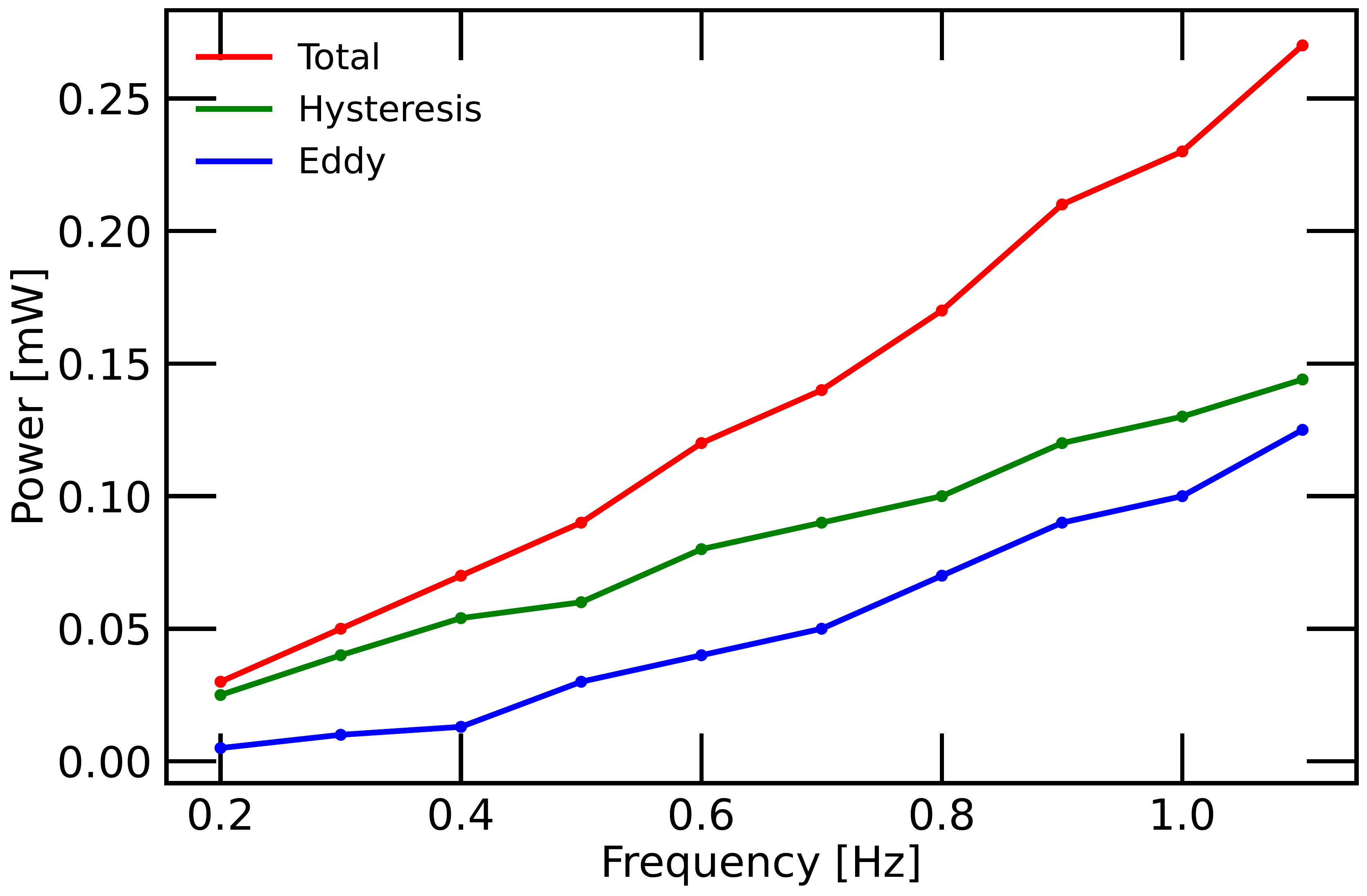}
    \end{tabular}
\end{center}
\caption[Spin down]{ \label{fig:spindown} 
Left: The spin-down measurements. The solid lines are experimental data, while the dashed lines are models. Right: The heat dissipation powers obtain from different rotational frequencies. Hysteresis losses are estimated by considering only the parameter $a_0$, i.e. setting $a_1=0$. A similar approach is used for eddy current losses by accounting only the contribution of $a_1$ term.}
\end{figure}

The left panel of Figure~\ref{fig:spindown} shows the spin-down measurements for a range of rotational frequencies from 0.2 to 1.1~Hz. We fit the rotational frequency data with the model Equation~\ref{eq:sol_spindown} to extract the parameters $a_0$ and $a_1$. The performance of the fit is shown in Figure~\ref{fig:spindown} (left). The typical values for $a_0$ and $a_1$ are $5.8\times10^{-3}$ $1/$s and $7.5 \times 10^{-4}$ \ 1/s$^2$. It is clear that the system is dominated by the hysteresis.

We assume the energy loss during the spin-down is dissipated as heat energy. The heat dissipation power $P$ is estimated as
\begin{equation}
    P = I\tau = I \dfrac{d \omega}{d t} \omega \\ 
      = I \left(  a_0 + 2 \pi a_1 f_{hwp} \right) 2 \pi f_{hwp},
\end{equation}
where $\tau$ is the torque of the rotor. The moment of inertia of the rotor is assumed to be $3.6\times10^{-3}$~kgm$^2$, and the corresponding heat dissipation is $ 0.23$~mW. 

The right panel of Figure~\ref{fig:spindown} shows the expected heat dissipation as a function frequency. The two contributions, hysteresis and the eddy current losses have a different frequency dependence. The hysteresis loss is higher than the eddy current loss at this rotational frequency rage.

\subsection{Modulated signal}
\begin{figure}[ht]
\begin{center}
    \begin{tabular}{c}
        \includegraphics[width=0.7\textwidth]{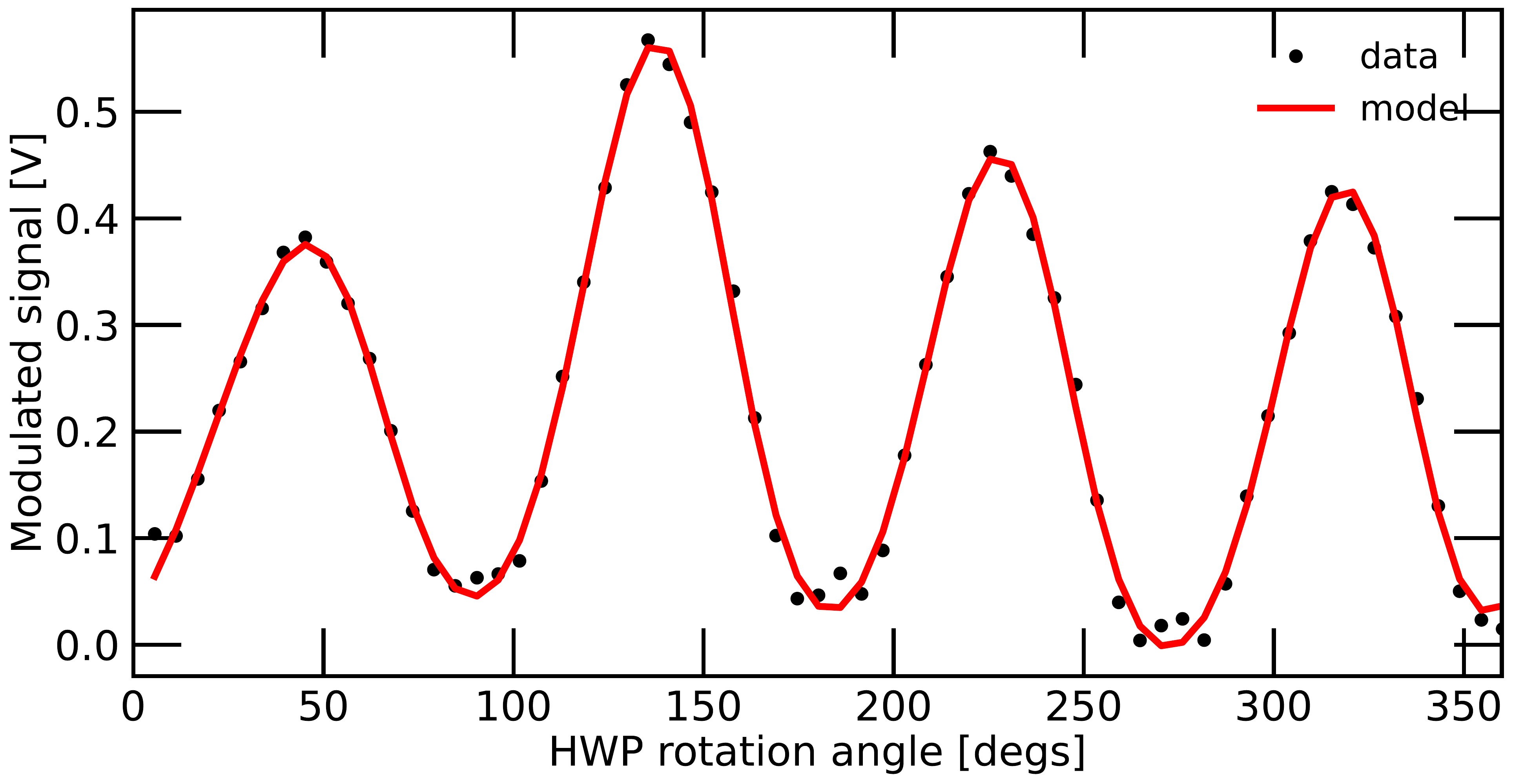}
    \end{tabular}
\end{center}
\caption[Modulated signal]{ \label{fig:mod_sig} The modulated signal at 90 GHz versus the HWP rotational angle. The black dotted curve is the experimental data. The red curve is the fitted model Equation~\ref{eq:modulate} compromised of $1f_{hwp}$, $2f_{hwp}$, $3f_{hwp}$ and $4f_{hwp}$ cosine waves. The seventeen free parameters for the fitting model are the amplitude and phase of each cosine wave and an overall offset.}
\end{figure}

Figure \ref{fig:mod_sig} shows the modulated signal as a function of the HWP angle for one revolution. We fitted the data with the model Equation~\ref{eq:modulate}. We first extract the modulation efficiency $A_4/A_0 \sim 0.94$ and the phase $\phi_4 = - 17.6$~degrees.
\begin{figure}[h]
\begin{center}
    \begin{tabular}{c}
        \includegraphics[width=0.5\textwidth]{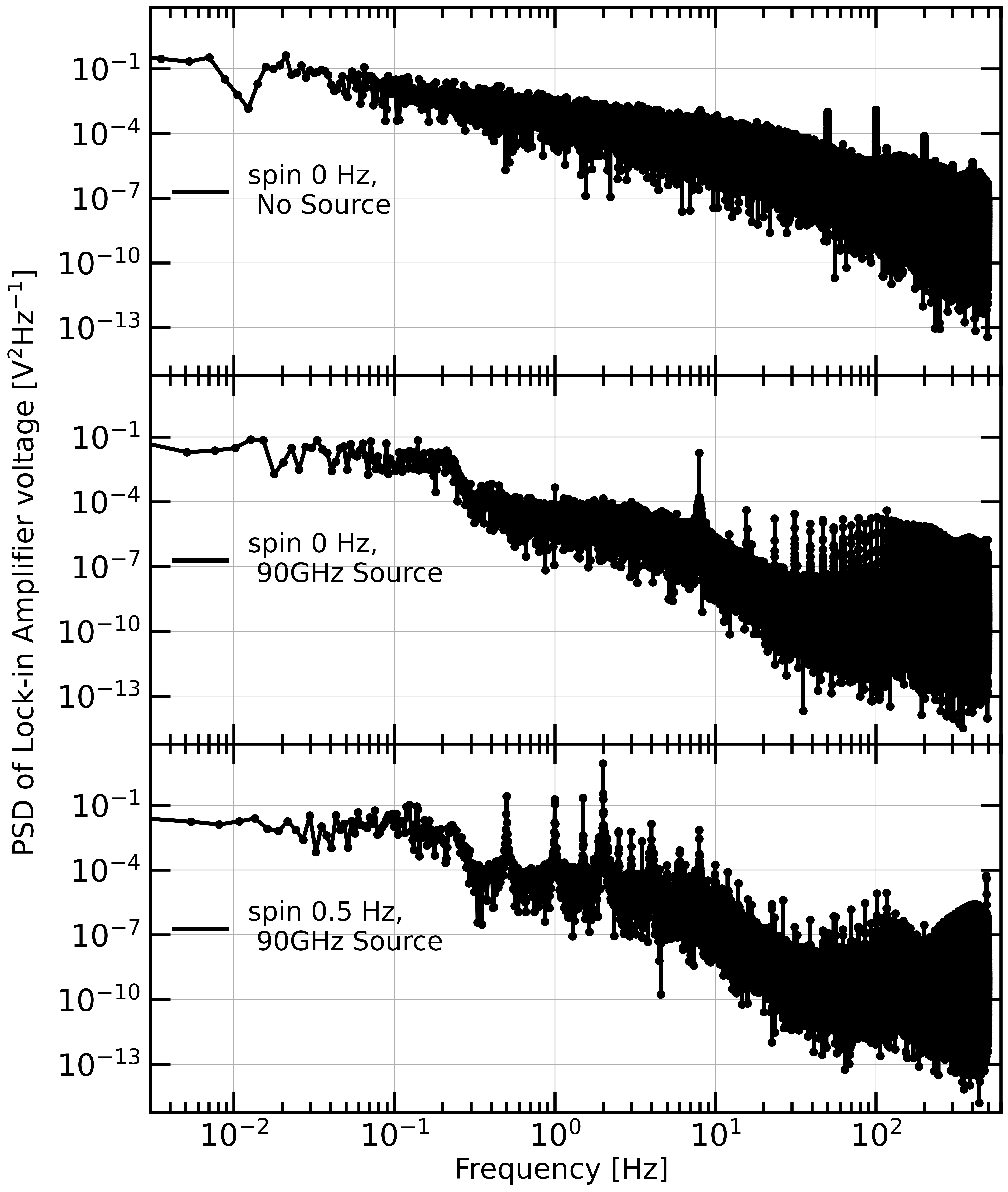}
        \includegraphics[width=0.5\textwidth]{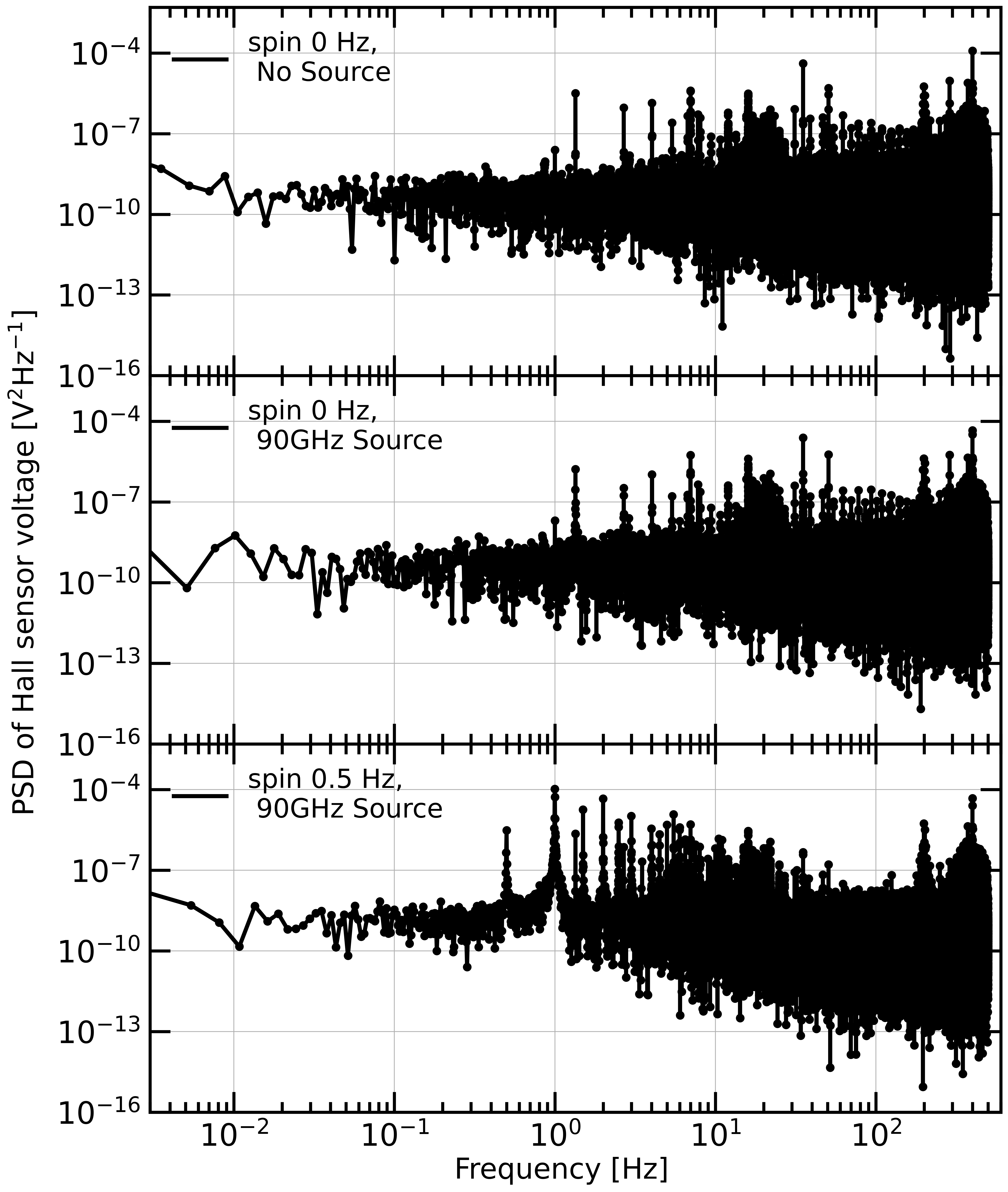}
    \end{tabular}
\end{center}
\caption[Lock-in amplifier]{ \label{fig:lockin} (left) The power spectral density of the modulated signal and (right) the Hall sensor signal for three different configurations.}
\end{figure}

Figure~\ref{fig:lockin} shows the PSD of the optical signal (left) from the lock-in amplifier and the PSD of the Hall sensor signal (right) for three configurations, no rotation with levitation and no input signal, no rotation with levitation and input signal, and rotation with levitation and input signal. We identified the peaks at the modulated frequency, $4f_{hwp}$, and rotational synchronous frequencies. The rich data set contains various subjects to be addressed. The examples are to identify the origin of all the peaks, determine the width and stability of the peaks, and demonstrate the demodulated PSD with the encoder data.

\section{DISCUSSIONS}
\label{sec:discus}
We have presented the current status of the prototype PMU system and the evaluation of the millimeter-wave polarimetric performance. We are aware that this system is much smaller than the prospective flight size. Thus, the results may not fully represent the effects we will encounter with the flight size. However, we plan to explore broad parameter spaces using this setup with the following motivations.

Our measurements in this paper are limited to a single electromagnetic frequency at 90~GHz. We can expand this to a broader frequency range and study the spectroscopic properties of the AHWP. We also plan to study the coupling between the AHWP and the neighbor geometry, e.g. baffle and aperture. Any rotational synchronous signal may originate from the AHWP itself or the rotational parts that hold the AHWP. In addition to relying on a standard optics simulation, it is beneficial to study this effect experimentally with fast turn-round design modifications. Also, it is of great interest to experimentally investigate the stability of the rotational synchronous signal.

This setup also helps prepare to combine with a TES detector and readout system. When the PMU operates with the TES and its readout electronics, there may be many additional effects, e.g. magnetic and EMI interferences and temperature of the AHWP itself, that we cannot investigate in this setup\cite{tommaso2020, shinya2022}. Therefore, the current setup is helpful in the hardware preparation and should also help disentangle some of the effects in future data with the TES and its readout system.

Last but not least, the small setup is easy to handle regarding the hardware preparation time and cryostat run time. Any measurement from this setup provides the recipe of the measurement methods and its sequence for the flight scale model.

\section{CONCLUSIONS}
\label{sec:conclu}

We have demonstrated a 1/10 scale prototype of the PMU for the \LB\ LFT. The prototype PMU AHWP consists of a five-layer AHWP with the anti-reflection sub-wavelength structure on the first and fifth layers. The rotational mechanism contains the SMB, the cryogenic holder mechanism,  and the encoder. We cooled down the prototype PMU to 10~K with a 4-K GM cryocooler. Then, we successfully levitated the rotor, and the AHWP rotated at several rotational frequencies. We carried out the spin-down measurement to estimate the heat dissipation due to the hysteresis loss and the eddy current loss. The coherent millimeter-wave source at 90~GHz is provided from outside of the cryostat to the inside, and the continuously rotating AHWP modulates the polarized signal. Given this setup presented in this paper, we are ready for more extensive tests, which are the preparation for the forthcoming full-scale model.

\acknowledgments 
We would like to thank Fabio Columbro and Peter Hargrave for fruitful comments. We thank the World Premier International Research Center Initiative (WPI), MEXT, Japan for support through Kavli IPMU. This work was also supported by JSPS KAKENHI Grant Numbers 18KK0083, and JSPS Core-to-Core Program number JPJSCCA20200003, A. Advanced Research Networks. \textit{LiteBIRD} (phase A) activities are supported by the following funding sources: ISAS/JAXA, MEXT, JSPS, KEK (Japan); CSA (Canada); CNES, CNRS, CEA (France); DFG (Germany); ASI, INFN, INAF (Italy); RCN (Norway); AEI (Spain); SNSA, SRC (Sweden); NASA, DOE (USA).
 
\bibliography{report} 
\bibliographystyle{spiebib} 

\end{document}